# Folksonomic Tag Clouds as an Aid to Content Indexing

Morgan Harvey, Mark Baillie, Ian Ruthven
University of Strathclyde
Department of Computer and Information Sciences
{morgan,mb,ir}@cis.strath.ac.uk

David Elsweiler
Friedrich-Alexander University
Institute of Computer Science
david.elsweiler@i8.informatik.uni-erlangen.de

## ABSTRACT
Social tagging systems have recently developed as a popular method of data organisation on the Internet. These systems allow users to organise their content in a way that makes sense to them, rather than forcing them to use a pre-determined and rigid set of categorisations. These "folksonomies" provide well populated sources of unstructured tags describing web resources which could potentially be used as semantic index terms for these resources. However getting people to agree on what tags best describe a resource is a difficult problem, therefore any feature which increases the consistency and stability of terms chosen would be extremely beneficial.

We investigate how the provision of a tag cloud, a weighted list of terms commonly used to assist in browsing a folksonomy, during the tagging process itself influences the tags produced and how difficult the user perceived the task to be. We show that illustrating the most popular tags to users assists in the tagging process and encourages a stable and consistent folksonomy to form.

## Categories and Subject Descriptors
H.5 [**Information Interfaces and Presentation**]: User Interfaces; H.3 [**Information Storage and Retrieval**]: Information Search and Retrieval

## 1. INTRODUCTION
Categorisation of resources on the Internet is a problem; the web is a constantly growing and evolving corpus of data. Search engines have been designed to provide automated indexing terms for these resources and generally work well for textual data. They are however less useful for more abstract items such as images or movie clips where human-generated keywords can be much more relevant. While the content of an image is generally obvious to its human owner, achieving automated pattern recognition accuracy is anything but trivial [1].

Social tagging systems have recently appeared as a popular method of providing human-generated categorisation data, particularly on the Internet. The popularity of such systems has led some to suggest that they may overcome traditional systems [2]. Unlike more traditional methods of categorisation (such as taxonomies or flat categories), these systems allow for collaborative and social development of the data set. By allowing users to organise data in a way that makes sense to them, rather than forcing them to use a predetermined - and potentially rigid - set of categorisations, the resulting folksonomies have the potential to be much more powerful, useful and relevant.

Tagging systems can be particularly effective when used to index photographs and sites such as Flickr have shown that manual indexing of images is extremely popular with most users and can result in some very useful categorisations. However issues can still arise as people tend to use different terms to classify the same resources.

In this paper we will examine some of the factors that can influence how people assign tags to a resource - in this case photographs - and examine the benefits of tag clouds - a feature of tagging systems that aim to improve the effectiveness of such systems. We discover that tag clouds can be useful as an aid when tagging resources and can reduce the perceived difficulty of assigning appropriate keywords to a photograph, in addition to increasing the measured consistency of the resulting tags. This results in potentially better index terms whilst still allowing for each individual users' more esoteric categorisations.

## 2. RELATED WORK
Many studies have shown that obtaining high consistency among different indexers is very difficult to achieve and can be affected by many factors including vocabulary, personal understanding of the resource and use of language [3]. It has been shown that indexers are more likely to agree on the concepts that should be indexed rather than on the terms that best represent the concepts themselves [4]. This lack of consistency is commonly a result of the "vocabulary problem" and [6] showed that the probability that two people describe a given object with a common word is less than 1 in 5.

The emergence of the web brings new problems: an ever-expanding and constantly evolving corpus generated by many users [7]. The sheer quantity of data involved makes it very difficult to manually index the collection [8]. Initially, portal services such as Yahoo! provided a useful way of accessing the early web's relatively limited content by providing a classification system based on a common shared vocabulary. However as the web has grown in size it is generally accepted that this method is no longer a feasible option. The de-facto standard on the web is now the fulltext search engine. Search engines index documents automatically by exploiting statistical methods, such as term frequency or link analysis, to establish keywords that describe resources and link structure as an indicator of an individual page's value. The result is an incomplete or inaccurate set of indexing terms [9], particularly when being applied to multimedia such as images [10].

[11] showed that folksonomy tags agree more closely with the human generated keywords than those that are automatically generated. It is possible therefore that folksonomies might offer a solution to this problem, providing a cheap source of semantically meaningful index terms.

Some pioneering work on the usage of folksonomies has been conducted [12] showing that while most people tend to tag for their own benefit, the categorisations they chose can be of use to the community as a whole. Further, other scholars have proposed that the community aspect of folksonomies helps to reinforce the tagging process and encourages further annotation [13], acting as a kind of feedback mechanism [15]. They fit in well with inherent "structure" of the Web, "the Web has an editor, it's everybody" [2] and underscore the transition of the web into a far more participatory medium. Tagging systems cover a diverse range of topics and resource types, from image tagging services such as Flickr to social bookmarking services like del.icio.us and CiteULike.

Despite its popularity, a number of issues have been identified with social tagging which are consistent with problems noted in all uncontrolled vocabularies [16], these issues can reduce the effectiveness of such systems for the indexing of resources. As with more traditional forms of indexing, increasing the consistency of index terms used to describe resources whilst allowing for personalisation of the folksonomy would provide a more useful system. In this paper we investigate how tag clouds (a common and popular visualisation of a set of weighted tags) might be utilised during the tagging process to reduce the negative impact of the issues noted above.

Tag clouds are "visual presentations of a set of words, typically a set of 'tags' selected by some rationale, in which attributes of the text such as size, weight, or colour are used to represent features, such as frequency, of the associated terms." [17]. Tag clouds can provide an overview of the overall theme (or gist) and content of the resource or collection being described.

bagpipes begging busker busking glencoe highlands hills kilt landscape mountain mountains piper scenery scenic scotland scottish snow sunny tartan winter

**Fig. 1. Example of a tag cloud with varying font size to indicate frequency.**

Studies [15] have shown the benefits of tag clouds in the retrieval process, particularly by allowing users to effectively browse through a collection. [18] showed that scanning the tag cloud requires less cognitive load than formulating specific query terms and that using the tag clouds to select keywords is 'easier' than thinking about what query terms will produce the best result. In this paper, we are interested in the use and effectiveness of tag clouds when used to assist in the tagging of resources. Do the tag clouds help to lower the difficulty of the tagging process and do they result in greater inter-indexer consistency that might yield better indexing terms, serving to reinforce the community aspect of tagging? Below we outline a study that addresses our aims. We examine how a broad-range of individuals tag photographs and analyse the influence that tag clouds have on the tagging process.

## 3. METHOD

To investigate these proposals, we conducted a web-based user study where participants were asked to assign keywords to a series of 12 images. The participants were asked to avoid using plurals and punctuation such as commas, exclamation marks or hyphens. They were also advised to keep their tags short, one or two word phrases rather than longer phrases or complete sentences. After tagging the images a short questionnaire was conducted to collect some demographic information. The photographs were chosen to represent a typical sample of images which are likely to be found on an online gallery such as Flickr, with some images having obvious themes and potential categorisations and others being more subjective. All participants were asked to tag the collection in the same order.

Two separate stages of the study were conducted in order to test our research questions:

The first stage was used to establish a folksonomy for the images in the study and to simulate a small and growing folksonomy. In this stage the tag clouds for each image were constantly changing with each participants' inputs being added to the existing cloud. For the second stage the clouds remained static and were based on the state of the folksonomy after the completion of the first stage.

Overall 79 people of varying age, level of education, employment positions, and degrees of computer literacy participated in the user study; 51 from the first stage and the remaining 28 from the second. The participants were not informed of purpose of the research and were not aware of that different participants were shown different information during the test.

**Table 1. Selected results of questionnaire.**

|  | Stage 1 | Stage 2 |
|---|---|---|
| **Median age range** | 19 - 29 | 19 - 29 |
| **Gender (Male/ Female)** | 41 / 10 | 20 / 8 |
| **Education (PG/UG/No Uni)** | 11 / 15 / 25 | 3 / 8 / 17 |
| **Computer use (Daily, most days)** | 42 / 9 | 22 / 6 |

In order to study how the provision of tag clouds influenced the keywords submitted, every second participant was presented with a tag cloud along with the image they were required to tag. The tag cloud was generated using the tags previous participants had assigned for that image. For the first stage of 51 participants, the tag cloud was different for each participant as all previous participants' tags were being added to the existing cloud. For the second stage the tag cloud remained static and was generated from the tags provided by the first stage. This cut-off was chosen as previous research has shown that folksonomies generally follow power-law distributions. Therefore the top tags by rank in a folksonomy tend to stabilise and remain constant after a sufficient number of users have tagged a given object [19].

The tag clouds were weighted by the normalised term frequency of tags used to describe the image by previous participants and only the 20 most frequently used tags up until that point were shown. Tags in the cloud were displayed in alphabetical order and tag frequency was mapped to tags in the cloud by varying the font-size. Participants who were shown the clouds were asked an extra question at the end of the study to ascertain whether they found the tag cloud useful or not. These participants were also instructed during the introduction on how to use the tag cloud and what it represented.

## 4. RESULTS

Several analyses were performed on both the tagging and questionnaire data. The following sections present our findings. Note that throughout the analyses the users are referred to in 4 separate groups:

- 1NC refers to participants from the first stage who were not shown the tag cloud
- 1TC is the participants from the first stage who were shown the cloud
- 2TC and 2NC are participants from the second stage again with and without the cloud

## 4.1 Tags

Participants submitted a total of 3467 tags, 723 of which were unique, the probability distributions for number of tags per user in all 4 groups were quite normal. The mean number of tags per participant for each group were as follows:

**Table 2. Tag count statistics**

|                     | 1NC   | 1TC   | 2NC  | 2TC  |
|---------------------|-------|-------|------|------|
| **Mean number of tags** | 39.8  | 46.2  | 37.9 | 43.9 |
| **Standard deviation**  | 29.93 | 22.37 | 29.2 | 24.7 |

Although a t-test showed that these results are not significant (p = 0.177, DF = 48 for 1NC vs. 1TC and p = 0.569, DF = 23 for 2NC vs. 2TC), they do indicate a trend towards more complete descriptions of the images which may be significant over a larger set of images and participants. It is also interesting that average number of tags for the 2 groups of participants over both time periods was so similar (1NC vs. 2NC and 1TC vs. 2TC) indicating that this trend may be consistent over larger populations. The standard deviations between the groups also indicate that the variance in tag counts submitted by users who were shown the coud is lower.

Analysis of the frequency of each tag used in the folksonomy shows the distinctive power-law distribution of tag frequency. That is a small number of tags are used very frequently with the larger majority only being used a small number of times. This tag frequency distribution has been identified in folksonomies in previous papers and is to be expected as they are an example of natural language use in a complex system. The mean number of times a given tag is used is 3.29 (95% CI [2.843 - 3.736]). Even though the folksonomy generated by the study is quite small and only covers a limited number of resources it still follows the expected power-law pattern seen in larger examples and indicates that the folksonomy may have reached a stable state.

## 4.2 Ease of tagging

In order to determine if tag clouds have an effect on perceived difficulty when tagging, each participant was asked to rate - from 1 (easy) to 5 (difficult) - how difficult they found it to come up with accurate tags. Each participant was also asked to rate their own organisational skill, again on a scale from 1 (poor) to 5 (excellent). The results are presented in the table below:

**Table 3. Tag difficulty and organisational skill for each group.**

|                             | Without cloud (1NC & 2NC) | With cloud (1TC & 2TC) |
|-----------------------------|---------------------------|------------------------|
| **Mean tag difficulty**     | 2.231                     | 1.8                    |
| **Mean organisational skill** | 3.41                    | 3.05                   |

A Mann-Whitney test showed statistically significant differences between the tag difficulties indicated by the participants in each group-pair (1NC & 2NC against 1TC & 2TC) (p = 0.03, 5% significance level), but did not show a significant difference in perceived organisational skill (p = 0.07, 5% significance level), therefore suggesting that the existence - or otherwise - of the cloud had an effect on the perceived difficulty of the tagging process.

This indicates that those participants who were shown the tag cloud were able to more easily complete the tagging process. This may be because the tag cloud reduced the cognitive-load on the participants by offering suggested keywords. Participants who were given the cloud were also asked to rate (from 1 to 5) how useful they found the tag cloud to be. The results from this question serve to back up the results discussed above as most participants found the tag cloud to be useful (mean 3.61; st. dev. 0.83) with only one participant stating that they did not find it useful at all.

## 4.3 Convergence and Group Consistency

In order to analyse how the use of the tag cloud affected participants' tagging behaviour a comparison was made between the tags a given participant used to describe an image and the group consensus (weighted list of the entire folksonomy) at that time. This metric was calculated by obtaining all tags a participant had submitted for a given image, then calculating the tag cloud for that image at the time the participant had tagged it. If the tag is present in the cloud list then the participant's "score" is increased by the weight that tag carried in the cloud. To normalise the result the final score for each user/image pair is divided by the user's tag count for that image. For the second stage of users (2NC and 2TC) the tag cloud remained static.

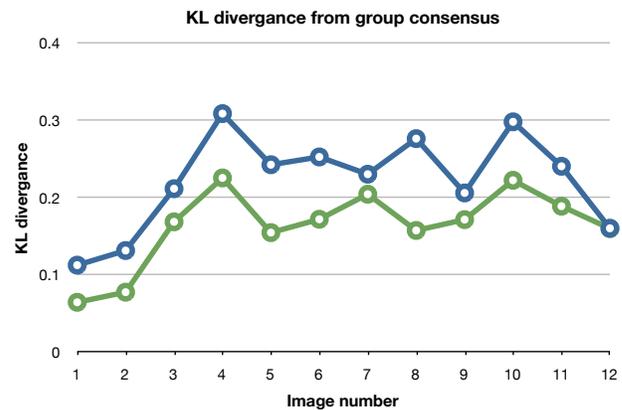

**Fig 2.** Graph of KL divergence from group consensus for users from both groups from stage 2. Green is TC2, blue is NC2.

We examined the tag distributions of both groups using Kullback-Leibler divergence; the difference between two (term) probablity distributions. The lower the KL score the closer the vocabluary distributions are. This analysis showed that for the first stage the difference between the two groups (i.e. 1NC against 1TC) was not statistically significant (p = 0.477), however it did indicate a consistent trend of the tag cloud group contributing more similar tags. With one exception, the tags contributed by the "with cloud" group diverged less from the group consensus. However, performing the same test on the tags submitted from the second stage (2NC against 2TC) does return a significant result (p = 0.018). A rank correlation test between terms used by both groups backed up the findings from the KL test with the 2 groups from the first stage tagging quite similarly (high correlation) and the 2 groups from the second stage tagging in quite a dissimilar fashion (low correlation).

It is interesting to note that while participants who were shown the cloud did perform better in this test (significantly so in the second stage) they still diverged somewhat from the group consensus. This suggests that they were not simply copying the existing tags in the cloud "wholesale," but instead were using the cloud to inform their own tag choices. This assumption is further backed up by qualitative data which were received from participants of the study. This data indicated that while participants found the cloud very useful, they only picked the terms from it that they agreed with and felt that this allowed them to think in more depth about the tags they chose. One user even commented that s/he felt "almost guilty" about choosing too many tags from the cloud and that made the tags s/he personally chose more descriptive and individual.

## 5. DISCUSSION AND FUTURE WORK

The results of the user study conducted support the notion that tag clouds may be useful as a mechanism for improving the tagging process. The existence of the tag cloud increased the volume and consistency of the tags entered by users and resulted in a more consistent and homogenous tag set for each image. Additionally the results of the questionnaire give strong credence to the notion that the tag cloud also decreased the difficulty of the task.

It is interesting that the data from the first stage do not result in a statically significant difference between the 2 groups, whereas the data from the second stage show a significant difference. This is because in the first run the continuously changing cloud resulted in the first few users only being shown a very small and incomplete cloud. In the second stage all users who are shown the cloud are presented with a more complete set of data which does not change over time.

The results of the data analyses show that the presence of a tag cloud of terms during the tagging process improves user performance, particularly for older, more stable folksonomies. Recent research [15] indicated that the most popular tags for a given resource (i.e. the ones given prominence in the tag clouds) tend to be the most relevant. However based on the data it would appear that this is not to the detriment of more idiosyncratic, personal tags as it is clear that users were not simply copying all of the tags from the cloud. Those shown the cloud were using the terms from the cloud they felt were most relevant in concert with tags of their own choosing. The study also shows that tag clouds reduced the participants' perceived difficulty of the task by a statistically significant margin and therefore (in terms of content indexing) reduces the negative impact of the translation stage on participant performance.

It is clear that future work in this area of research could be of benefit and that further analysis of the data obtained in the study may yield additional results of interest. It may be useful to investigate, for example, how the use of tag clouds as a tagging aid in concert with automated tag suggestion methods might result in a more useful and self-reinforcing tagging system, even when the number of existing tags available for a given resource is small. I am currently using the results of this study to assist in the development of statistical models of folksonomic systems and browsing and searching interfaces that make use of these models.